**PROTOPLANETARY MIGRATION AND CREATION OF SCATTERED PLANETESIMAL DISKS.**
Bruce D. Lindsay and Truell W. Hyde, CASPER (Center for Astrophysics, Space Physics and Engineering Research), P. O. Box 97316, Baylor University, Waco, TX 76798-7316, USA, phone: 254-710-2511 (e-mail: Truell_Hyde@baylor.edu.)

**Introduction:** The creation and evolution of the region of planetesimals known as the Kuiper Belt has recently become an important topic of study in the field of planetary formation. The Kuiper Belt is made up of a very large number of planetesimals that orbit the Sun at distances beyond Neptune, perhaps even as far away as 1,000 AU [1]. These planetesimals are believed to either be relics of the planetary formation process that originated in this transNeptunian region and avoided subsequent close encounters, or planetesimals that were created closer to the central star only to be scattered outward by larger protoplanets. The stability of the planetesimals in the Kuiper Belt is governed primarily by the gravitational influence (or lack thereof) of Neptune. It has been shown that most bodies in the Kuiper Belt with high orbital eccentricities are found in mean motion resonances with Neptune. Planetesimals with modest eccentricities and high inclinations (resulting in orbits that are stable over periods comparable to the age of the planetary system) are able to avoid being perturbed out of their orbits since their perihelia are well separated from Neptune [2].

It has recently been suggested that scattering mechanisms created by mutual gravitational interactions could cause inward migration by protoplanets as well as the formation of planetary system features such as the Kuiper Belt [3]. This could explain the discovery of extrasolar hot Jupiters, which are currently thought to have formed in larger orbits and then moved closer to their parent stars via some migration mechanism [4]. By numerically modeling the effects of one or more growing protoplanets on a ring of planetesimals, it should be possible to determine the point in its evolution at which it had the most influence on the protoplanetary system, scattering bodies outward and creating structures similar to the Kuiper Belt. Conversely, this would also correspond to the time period during which the evolving protoplanet would be able to move inward due to such interactions.

**Simulation Model:** The numerical model employed by this study calculated planetesimal and protoplanet trajectories using a fifth-order Runge-Kutta algorithm. Individual trajectories were determined considering the mutual gravitational forces between the planetesimals and protoplanet. The temporal evolution of an annulus of planetesimals and a larger protoplanet at one of its edges was examined. The initial planetesimal velocity distribution was defined as Keplerian.

In the first six simulations, 500 planetesimals with varying masses were examined, beginning in circular orbits with their semimajor axes scattered randomly between 10 and 20 AU. Orbital inclinations were also randomly selected. The planetesimal masses and maximum orbital inclinations are given in Table 1. The simulations were run to model the evolution over a period of ten thousand years.

Table 1. Initial conditions for Simulations 1-6

| Run # | Planetesimal Mass (Earth masses) | Maximum Inclination (radians) |
|---|---|---|
| 1 | 0.02 | 0.01 |
| 2 | 0.05 | 0.01 |
| 3 | 0.1 | 0.01 |
| 4 | 0.1 | 0.05 |
| 5 | 0.1 | 0.05 |
| 6 | 0.05 | 0.03 |

In the other four simulations, one thousand planetesimals were assumed, beginning in circular, coplanar (inclination equal to zero) orbits with their semimajor axes scattered randomly between 15 and 25 AU. Each planetesimal was given a mass one-tenth that of the mass of the Earth. A protoplanet with a varying mass was located at the outer edge of the planetesimal annulus (25 AU) as given in Table 2. The protoplanet's mass was varied to determine the critical mass for disk scattering and its relationship to both planetesimal and protoplanet migration. This is of vital importance since this should determine when the growing protoplanet has its maximum impact on the evolution of the protoplanetary system. These simulations examined the system's evolution over a span of five thousand years.

**Results and Conclusions:** In simulations with a protoplanet located at the outer edge of the planetesimal ring, more planetesimals than in the case without the protoplanet were scattered both inward and outward from the initial limits of the ring. As can be seen in Fig. 1, the average eccentricities for the planetesimals scattered inward are generally smaller than the minimum value (or stability limit)

PROTOPLANET MIGRATION AND SCATTERED PLANETESIMAL DISKS: B. Lindsay and T. Hyde

for them to undergo subsequent close encounters with the protoplanet. The stability limit corresponds to the eccentricity required for the aphelion (if the planetesimal is closer to the star than the protoplanet) or the perihelion (if the planetesimal is farther away from the star than the protoplanet) to allow the planetesimal s orbit to cross that of the protoplanet. This infers that gravitational interactions between the planetesimals themselves are a possible cause of this inward scattering.

Table 2. Protoplanet Mass

| Run # | Protoplanet mass (Earth masses) |
|---|---|
| 7 | 200 |
| 8 | 100 |
| 9 | 50 |
| 10 | 25 |

Conversely, the average eccentricities for outwardly scattered planetesimals are well above the minimum value needed for close protoplanet-planetesimal encounters to occur. This suggests that such events in the past may have been responsible for outward scattering events. The outward flow of planetesimals corresponds to an inward migration of the protoplanet, as can be seen in Tables 3 and 4. It can also be seen that the number of inelastic collisions between planetesimals increases when the protoplanet mass decreases.

Table 4 gives the results of the interactions between the planetesimals during each simulation. All collisions were found to be inelastic, although the algorithm was equipped to resolve elastic collisions. The data shows that the presence of a single larger object in the vicinity of a ring of planetesimals produces an instability in the disk. This causes planetesimals to be scattered out of the region (or the protoplanetary system altogether) at a much faster rate than if there were no larger objects to disturb the planetesimal disk.

Table 3. Final Semimajor Axes for the Protoplanet

| Run # | Final Semimajor Axis Of Protoplanet (AU) |
|---|---|
| 7 | 23.8554 |
| 8 | 24.3806 |
| 9 | 19.8388 |
| 10 | 23.7027 |

Table 4. Planetesimal Interaction Results

| Run # | Collisions | Escapes | Scattered Inward | Scattered Outward |
|---|---|---|---|---|
| 1 | 0 | 0 | 5 | 12 |
| 2 | 0 | 0 | 6 | 19 |
| 3 | 0 | 4 | 13 | 20 |
| 4 | 0 | 0 | 7 | 19 |
| 5 | 0 | 0 | 4 | 27 |
| 6 | 0 | 0 | 3 | 10 |
| 7 | 2 | 90 | 93 | 171 |
| 8 | 3 | 41 | 53 | 238 |
| 9 | 4 | 114 | 134 | 292 |
| 10 | 5 | 70 | 79 | 191 |

Figure 1. Average Eccentricity Distribution for Planetesimal Ring with One Protoplanet at Outer Edge (Simulation 10)

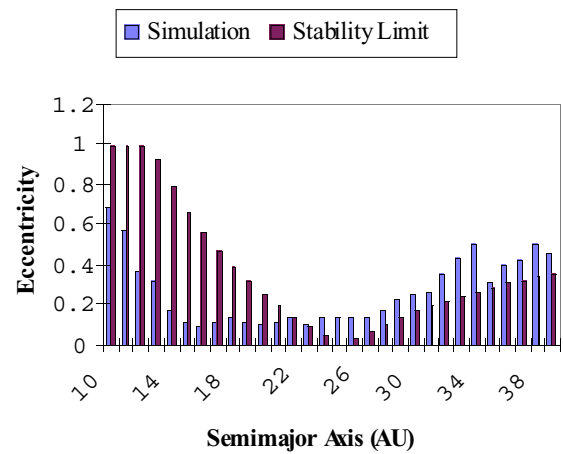